# Physics Computational Literacy: An Exploratory Case Study Using Computational Essays

**Tor Ole B. Odden[1], Elise Lockwood[2], and Marcos D. Caballero[1,3]**
**[1]Center for Computing in Science Education, University of Oslo, Oslo, Norway**
**[2]Department of Mathematics, Oregon State University, Corvallis, OR**
**[3]Department of Physics and Astronomy & CREATE for STEM Institute, Michigan State University, East Lansing, MI**

## ABSTRACT

Computation is becoming an increasingly important part of physics education. However, there are currently few theories of learning that can be used to help explain and predict the unique challenges and affordances associated with computation in physics. In this study, we adapt the existing theory of computational literacy, which posits that computational learning can be divided into material, cognitive, and social aspects, to the context of undergraduate physics. Based on an exploratory study of undergraduate physics computational literacy, using a newly-developed teaching tool known as a computational essay, we have identified a variety of student practices, knowledge, and beliefs across these three aspects of computational literacy. We illustrate these categories with data collected from students who engaged in an initial implementation of computational essays in a 3rd-semester electricity and magnetism class. We conclude by arguing that this framework can be used to theoretically diagnose student difficulties with computation, distinguish educational approaches that focus on material vs. cognitive aspects of computational literacy, and highlight the benefits and limitations of open-ended projects like computational essays to student learning.

## I. INTRODUCTION: NEED FOR A THEORY OF COMPUTATIONAL LEARNING IN PHYSICS

The field of physics is becoming increasingly computational. Within the last 30 years, computation has grown in status and sophistication, to the point that it is now regarded by many as a third pillar of physics, on equal footing with theory and experiment [1,2]. A major goal in physics education research is to make physics education more authentic to the discipline, and based on these trends many physics programs will soon need to tackle the challenge of robustly integrating computation into their curricula [3].

We contend that physics is uniquely suited to helping students understand computation, and computation is uniquely suited to helping students understand physics. The history of computational physics education bears this out: for example, in the late 1960s researchers at MIT developed the LOGO computing language in part to help children explore physics in a fully Newtonian "microworld" [4]. Later, in the 1980s, Andrea diSessa developed the BOXER system, which was successfully used to teach certain physics and calculus concepts to students as young as 6th grade [5,6]. More recently, there has been an explosion in the variety of different games, apps, and simulations that leverage computation to help students learn physics, such as PhET simulations [7,8], Physlets [9,10], Tychos Hackable Simulations [11], and exercises posted to the Online PICUP Repository [12]. At the undergraduate level there are also several curricula that leverage the visual and modeling-based possibilities in programming [13,14].

Despite this progress, computation is not yet widely integrated into physics teaching. Apart from a few well-established curricula like Matter & Interactions and certain mathematical tools (e.g., Mathematica and MATLAB), most university-level physics courses have relatively little engagement with computation [15–17]. There are numerous systemic factors related to this lack of adoption [18], but we contend that these factors alone should not deter the physics education research community from exploring ways in which computation can positively influence physics teaching, and vice versa. Thus, we see a need for the development of theories of teaching and learning that can address the unique challenges and affordances that computation brings to the table and that can help guide the implementation of computation across courses and curricula.

In this paper, we propose one such theory, which we call *physics computational literacy.* Our goal is to draw on the general theory of computational literacy proposed by diSessa [5,6] to articulate a framework for computational learning that is specific to physics. We have developed this theory using a newly-developed teaching tool designed to bring out different elements of computational literacy, known as a computational essay. In what follows, we describe and elaborate our theory by presenting excerpts from a case-study of the use of computational essays in a university setting, and we use this case study to argue for the ways in which physics computational literacy could be a useful framework when integrating computing into physics classrooms and programs.

## II. THEORETICAL FRAMEWORK: COMPUTATIONAL LITERACY

### A. A theoretical description of computational literacy and its application to physics

For this study we are drawing on the theory of *computational literacy*, originally proposed by Andrea diSessa [5,6]. In his seminal book "Changing Minds: Computers, Learning, and Literacy," diSessa argues that computation is rapidly becoming a new literacy, at the same level of importance as mathematics, reading, and writing. In other words, computation is becoming increasingly common and necessary for everyday life and professional practice, while simultaneously making possible a new set of skills and ways of thinking. Although this argument has been articulated by numerous other authors (e.g., [19–21]), diSessa goes one step further by arguing that the ways we use computation is structurally similar to the ways we use print and mathematics: each set of skills is based on a specific representational system (code, print, or numbers) that have certain rules for its use (syntax, algebra, grammar) and dedicated intellectual purposes (programming, communication, calculation).

Based on this argument, diSessa proposes that we can understand the essential elements of computational literacy by analogy to the core elements of print or mathematical literacy. In particular, diSessa [5] argues that there are three "pillars" of computational literacy (p. 6-9). First, there is the *material pillar*, which consists of familiarity and fluency with the basic representational system underlying all programming, computer code. Such familiarity is a necessary precursor to computational literacy in the same way that one must be familiar with letters and sentences, or numbers and mathematical symbols, to be print or mathematically literate. In order to acquire this material fluency, one must learn to program, at least at a basic level, including operations like assigning variables, defining functions, and running simple scripts. One must also be familiar with at least some of the structural components of code such as syntax, objects, and libraries, and the tools necessary to program such as integrated development environments.

The second pillar of computational literacy is what diSessa calls the *cognitive pillar*, which consists of the ways in which we can use this material basis to augment our ways of thinking and improve our understanding of the world. In the same way that humans use mathematics to simplify or to add precision to tasks, from engineering and architecture to grocery shopping and everyday decision-making, computation can be used to expand the space of tractable problems and broaden ways in which we acquire new knowledge. Thus, this pillar is "cognitive" in that it extends our cognition, allowing us to think about and understand the world in new ways. Scientists, for example, use computation to store and parse large amounts of data that would be intractable to do by hand; create visual representations that would be difficult to generate in other ways; make predictions about the future behavior of systems; and send and receive huge amounts of data across large distances. Each of these applications allows for new insights and ways of acquiring knowledge that would be difficult or impossible without computation.

Acquiring cognitive computational literacy therefore involves learning a new set of skills beyond the fundamentals of programming—namely, the ways in which one can apply computation to tasks. In other literacies, one might learn different writing styles (argumentative essays, news reports, technical summaries) or different applications of mathematics (measuring quantities, solving equations, statistical analysis). In the same way, in computational literacy one must learn ways to apply the fundamental tools of computation to real-world problems and situations.

Computation is never done in a vacuum—one is always programming with others, whether through collaboration on projects, consulting documentation, or building on others' code. With this in mind, the third pillar, which diSessa calls the *social pillar*, focuses on the ways in which one communicates with and about computation within a community. In the same way that both reading/writing and mathematics are used to communicate with others (as well as create, store, or calculate new information for oneself), diSessa argues that computation will always have an inherent social dimension that must be taken into account in any robust theory of computational learning.

To summarize, according to this theory, there are three aspects of literacy that diSessa argues one must cultivate in order to become computationally literate.

- **Material Computational Literacy:** the mechanics, techniques, and knowledge involved in the act of programming. A novice grasp of this material literacy would involve learning the basic syntax of at least one programming language, including subskills like assigning variables, creating loops and functions, as well as the techniques of running and debugging simple scripts. At higher levels of material computational literacy, one would be able to program in multiple languages, define classes and objects, and understand the inner workings of different types of computers and operating systems.

- **Cognitive Computational Literacy:** the ways in which we apply programming to extend our thinking and understand the world. This aspect of computational literacy is focused on applying programming to different problems and contexts in order to accomplish tasks that one would be unable to do without computation. At a novice level, this would involve using programming to solve simple problems such as predicting the outcome of random events, the motion of an object under certain forces, or the behavior of a predator-prey systems. At more advanced levels, it might involve applications like simulating the dynamics of a many-body system to predict its behavior over time, or generating field and contour plots to visualize a complex potential.

- **Social Computational Literacy:** the ways in which we communicate with and about computation to other people. This includes everything from the communication practices within project teams to how one structures one's code to make it more readable. At a novice level, this might involve common practices such as commenting code, explaining the meaning of one's code to others, and writing simple reports on computational projects. At a more advanced level it could involve presenting computational work to a research group, both verbally and through the use of an annotated notebook, as well as consulting and contributing to code documentation.

Although we have presented these three pillars as separate, in practice they are almost always entangled with one another. That is, nearly every application of computer programming will require some amount of material, cognitive, and social computational literacy on the part of the programmer. Indeed, one will seldom see students cultivating them separately from one another, other than in certain decontextualized environments like basic programming courses or coding "bootcamps." There is even more overlap when students learn to program within a particular disciplinary context such as the Matter & Interactions curriculum or P$^3$ learning environment [13,14].

However, we see it as theoretically productive to try to disentangle these three pillars of computational literacy, however artificial this may be in practice. At an individual level, distinguishing between material and cognitive computational literacy can allow us to diagnose student difficulties—for example, is a particular student struggling with their simulation because they are having difficulty connecting their knowledge of forces to their code (cognitive) or because they haven't yet understood the ways in which variable assignation works (material)? At a broader level, these distinctions may allow us to determine the strengths and weaknesses of different educational approaches. For example, some approaches require students to take a basic programming course in a computer science department before they use computation in physics; such approaches could be characterized as emphasizing the material aspects independent of the cognitive. Those that introduce computation within a physics course, teaching students only the bare minimum amount of coding necessary to modify and run a simulation, could be characterized as prioritizing cognitive computational literacy over the material aspects. And because it has not yet been a focus of the literature, many educational approaches are likely to ignore the social aspects of computational literacy. As the field grows in its attention to the integration of computing into STEM (and physics in particular), it will be useful to have language to articulate such phenomena and distinctions among individuals and across programs more broadly.

Based on this general theory, we conceptualize physics computational literacy as existing at the intersection of the domains of physics, computation, and the more general social sphere, with each domain contributing to one of the pillars described above. We illustrate this theoretical overlap in Figure 1. The general domain of computation (encompassing computational practices, knowledge, and beliefs) contributes the material aspects of physics computational literacy. The domain of physics (encompassing common practices, knowledge, and beliefs students acquire while studying physics) contributes the cognitive elements—applications of programming for learning and discovery. And the more general social domain (encompassing the ways in which students collaborate and communicate across disciplines) contributes the social aspects. With this conceptualization, one can begin to imagine how to analytically disentangle the three pillars of physics computational literacy. In particular, elements of students' computational activities that are primarily focused on their interaction with the code, and which they might use across a variety of different disciplines, contexts, and problems, would be classified under the material pillar. Elements that are focused on ways students use computation to solve problems and gain new insights about the world, regardless of their particular implementations in code, would be classified under the cognitive pillar. And elements that focus on communication and collaboration with others would be classified under the social pillar.

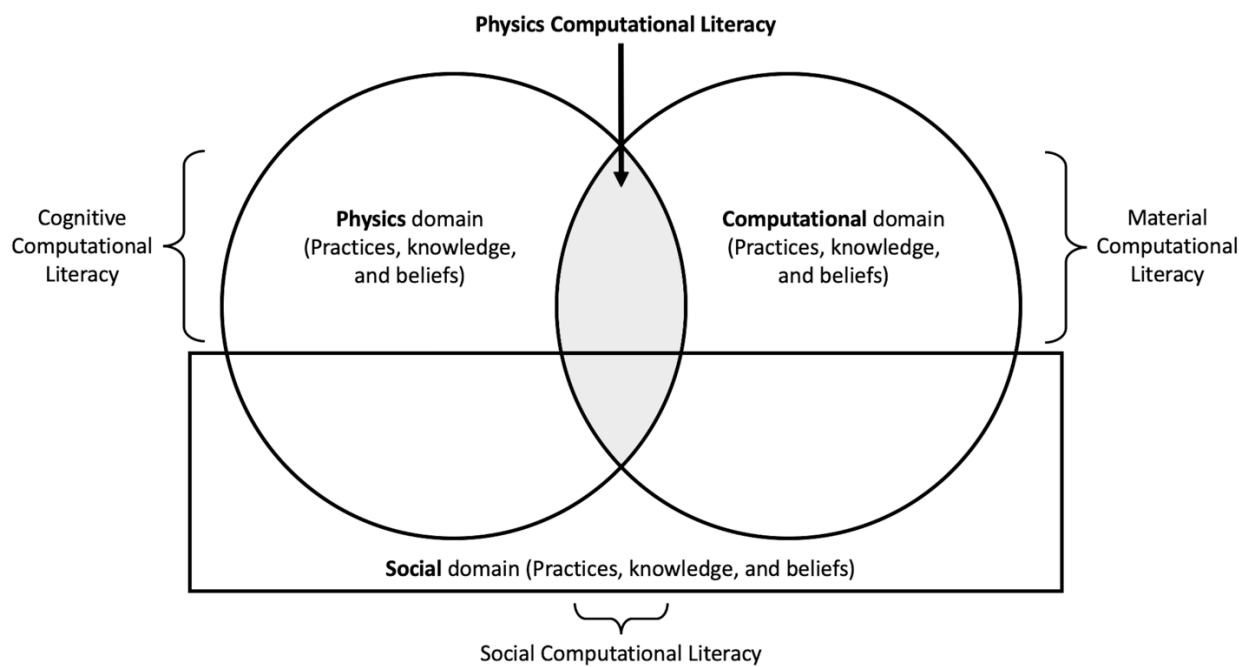

**Figure 1:** Theoretical positioning of physics computational literacy relative to physics, computational, and social domains.

### B. Computational literacy's relation to socio-cognitive theories and other theories of computational learning

Some readers may notice similarities between this theory and other socio-cognitive theories like activity theory [22,23] and distributed cognition [24,25]. Although diSessa does not cite these theories directly, there is a rich tradition of applying both activity theory [26] and distributed cognition [27,28] to the study of human-computer interactions, especially during the period when diSessa proposed his theory of computational literacy. So, we infer that these theories provided at least some inspiration for the computational literacy framework. However, we feel that the advantage of the computational literacy framework is that it is significantly more applied than either activity theory or distributed cognition. That is, both of these theories provide large-scale frameworks, which are useful for modeling the ways in which humans interact with tools and their environments but are less useful in helping guide educational design and practice. Computational literacy, in contrast, is explicitly meant to help guide one in thinking through both how to help students become more fluent and sophisticated in their use of computation, and how that use will affect their learning of other topics, subjects, and disciplines. Because this ultimately aligns with our motivation and interest, we feel that computational literacy is an appropriate perspective to adopt.

We would also like to distinguish between the theory of computational literacy and related theories, such as digital literacy [21] and computational thinking (e.g., [19,29,30]). Digital literacy is a general term referring to skill or ability to use modern digital devices such as smartphones, computers, software applications, and internet browsers. Though certainly important, these skills have little relation to computational literacy as defined above. It is quite possible (and, we would argue, common) to be quite digitally literate but have little-to-no computational literacy. That is, while many people are able to use their smartphones and computers for everyday tasks (e.g., checking email, reading the news), it is much less common to have the skills and confidence to be able to open a developer environment and write a script to solve problems or accomplish tasks.

Computational thinking generally refers to the ability to think in computational, procedural, or algorithmic terms for problem-solving [19,29,31,32]. Although this type of problem-solving is certainly an important part of computational literacy (especially within the cognitive pillar), researchers who focus on computational thinking often go to great lengths to divorce computational thinking from the act of programming, as exemplified by the movement towards "unplugged" computational activities [33]. In contrast, computational literacy explicitly incorporates use of programming in the material pillar. Computational literacy also takes into account the social aspects of computation, which few definitions or characterizations of computational thinking do. Additionally, computational thinking, in part due to the popularity of the theory, has been criticized as becoming an over-broad "higher order thinking skill," similar to other skills like "critical thinking" [6]. For these reasons, we see computational literacy as a more focused theoretical lens with which to study the intersection between computation and physics in education.

Since its proposal, the theory of computational literacy has gained some limited use in studies of undergraduate engineering education [34], the maker movement [35], and the broader field of literacy studies [20]. Magana et al. [34], in particular, extended the theory of computational literacy to incorporate the discipline-specific aspects of engineering education, merging it with a so-called "practice perspective" that focuses on learning as participation in disciplinarily authentic contexts using tools authentic to the discipline. In many ways, this is a natural next step for the theory. One of diSessa's arguments is that each community that adopts a particular literacy ends up developing their own "literatures"—ways of using the tools of that literacy. Becoming "literate" therefore involves becoming versed in the ways of using that tool. This phenomenon has long been studied in the field of the learning sciences, which has shown that different communities use tools such as mathematics in different ways—basic mathematical practices are used differently by, for example, grocery shoppers [36] versus tailors [37]. In the same way, a physicist will almost certainly use programming in a different way than a commercial software engineer, data scientist, or web designer, and consequently will need training in a specific subset of material, cognitive, and social aspects of computation.

Furthermore, these ways of using computation are likely to change and develop over a students' educational career, so that an undergraduate physics major will almost certainly use programming differently than a graduate student or professional researcher. In this regard, becoming computationally literate is one step on the road to becoming an expert in a computational discipline [38–40], and it can be argued that the framework of computational literacy uses an "expert-like" lens to evaluate students' progress towards that goal. However, in this study we are constraining our focus to the computational literacy of relative novices, undergraduate physics majors, who are on their way to disciplinary expertise. That is, we are attempting to flesh out diSessa's general theoretical description of computational literacy by defining the specific material, cognitive, and social practices that collectively form the computational literacy of undergraduate physics students.

We wish to emphasize again that our goal with the study is exploratory, and we are not aiming to offer complete theoretical coverage of all possible aspects of computational literacy in physics; in fact, we feel this would be both methodologically intractable and not particularly useful to the field because different physics programs will likely be implementing computation to different degrees. Instead, we are aiming to adapt diSessa's framework to undergraduate physics, so that it can be used to think through the different aspects of computation that teachers and curriculum designers may want to incorporate. This framework might then allow us to concretize the goals for computational education in physics, adding theoretical consistency to the current effort to develop standards for integrating computation into physics programs [3]. For this reason, in this study our primary research questions are as follows:

1. What are essential components of physics computational literacy at an undergraduate level[1], and how do they manifest in practice?

2. How do we support physics students' development of computational literacy in an undergraduate physics setting?

### III. CONTEXT AND METHODS: COMPUTATIONAL ESSAY DEVELOPMENT AND USE AT THE UNIVERSITY OF OSLO

To address these research questions, we situated our study within a setting in which computation is well integrated throughout the physics curriculum, the physics department of the University of Oslo, Norway. Since 2003, computation has been a cornerstone of the University of Oslo's physics program. Physics majors at the University of Oslo do not take any physics during their first semester; instead, they take a programming course, a calculus course, and a numerical methods course that teaches basic computational techniques for solving mathematical and scientific problems. Once they begin taking physics courses (in their second semester), students use this programming foundation to write simulations as part of their weekly homework assignments and exams. For example, students frequently simulate the motion of various objects in their 2nd-semester mechanics course, and are exposed to more advanced techniques like the method of relaxation and finite difference methods during their 3$^{rd}$ semester electricity and magnetism course, which spans introductory concepts through the first seven chapters of Griffiths' *Introduction to Electrodynamics* [41]. The department also hosts a robust computational physics research program, with numerous possibilities for undergraduate research. This setting therefore presents a unique opportunity to investigate the nature of undergraduate physics computational literacy in practice, because students spend their first semester developing their material computational literacy through their programming courses, and only later begin to focus on the cognitive aspects in their physics courses. Social computational literacy, as far as we know, has not yet been explicitly emphasized in any courses.

---

[1] We acknowledge that some of these ideas may be applicable more broadly to other disciplines besides physics, and to other populations other than physics undergraduates specifically. However, we feel comfortable restricting our claims to this population, and we discuss potential implications for other fields in the *Theoretical implications of the computational literacy framework* section.

In 2018, we began development of a new type of teaching tool intended both to support students' development of physics computational literacy and to allow us to study that development. We called this tool a *computational essay*. Computational essays were originally proposed by diSessa [5] as a form of writing that uses text, along with small programs, interactive diagrams, and computational tools to express an idea. The same concept was more recently picked up by Stephen Wolfram, the chief designer of Mathematica, who conceptualized computational essays as documents that use text, computer input, and computer output to explore and communicate ideas [42]. Computational essays are, in short, a type of essay that explicitly incorporate code to support their theses. They include all of the elements one would expect in an ordinary essay: an introduction, thesis statement, body paragraphs, and conclusion. They also have a similar set of goals: to present a step-by-step argument or explanation. However, the argument in a computational essay is driven by the output of various blocks of code, with the text serving both to explicate the meaning of the code and to explain the output.

**Narrative text**                                    **Computer Code**

Title and Introduction

Pictures and Diagrams

Pictures and Diagrams

Importing packages

Model parameters

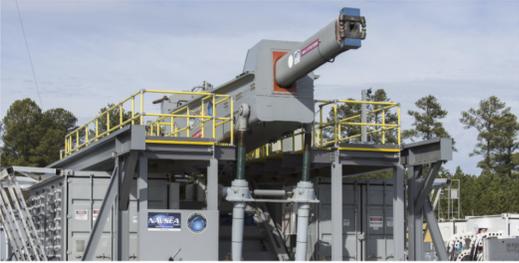

**Figure 2:** Example computational essay, showing the mix of text, code, and pictures. Example essays and student essays are both available at
https://uio-ccse.github.io/computational-essay-showroom

In order to achieve this integration of code and text, computational essays must be written in development environments that allow users to mix the two. One of the most common examples of these environments are so-called programming *notebooks.* Notebooks are programming environments that have a series of blanks (called "inputs") into which users can type both code and text. Inputs can hold as much code or text as the user desires, from single lines to whole programs or paragraphs. Particular code inputs can be run multiple times independent of the rest of the notebook, allowing users to make and test small-scale changes in a program without re-running the entire script. There are several types of notebook software currently available, including the widely-used, proprietary Mathematica notebook and the newer Jupyter notebooks [43], which are open-source environments that allow users this same flexibility with Python and other commonly-used programming languages.

Notebooks are becoming increasingly common tools for data scientists and professional physicists, who use them for both exploratory analysis and presenting findings [44,45]. However, to our knowledge notebooks and computational essays have not yet been widely used in educational environments. This, we feel, is a missed opportunity, because computational essays have the potential to address all three pillars of computational literacy: students are required to write code (material), apply it to a particular problem (cognitive), and explain that code in text and visuals (social). So, in the fall semester of 2018 we carried out an exploratory study to try out computational essays in the University of Oslo's 3rd-semester undergraduate electricity and magnetism course for physics and engineering majors.

For this exploratory study, the students in the course were given the option to write a computational essay as an alternative to a mandatory presentation-based project that all students had to complete. Those who chose to participate were challenged to conduct an open-ended computational investigation of some phenomenon related to the course content, then to write a computational essay that summarized their investigation. They were given suggestions for problems and topics to pursue, but were encouraged to work on topics that they found interesting. Students were allowed to work individually or in pairs, and they were given approximately 4-6 weeks to work on their essays.

To help scaffold the project, we provided students with a project description outlining our expectations, and an example computational essay on the topic "how much current would a railgun need to launch a resupply package up to the International Space Station?" (shown in Figure 2). We also provided the students with several basic simulations of electricity and magnetism phenomena, written in Jupyter notebooks, including a simulation of a cyclotron, storm cloud, lightning strike, and magnetic trap [46]. These simulations were meant to act as "seeds" for the student projects, in that students were encouraged to augment and build off of them as they were performing their analyses. The simulations spanned a variety of standard computational physics methods, including numerical superposition, Euler-Cromer integration, and the method of relaxation. Each simulation also included several optional prompts and questions that the students could use as inspiration if they were having difficulty coming up with ideas themselves. During the time when they were working on the computational essays the first author additionally staffed a special "helpdesk hour" for 2 hours twice a week to answer questions and help students with issues related to programming, physics, or project expectations. At the end of the semester, students presented their essays to their peers in mock "research-group meetings," and the resulting essays were graded pass/fail.

In total, 17 students participated in this initial implementation of computational essays, working singly or in pairs to produce a total of 11 completed essays. Three of the participating students were female and 14 were male; all female students worked in a pair with a male partner. Thirteen of the students were physics majors, two were physics pre-service teachers, and two were in a related major of materials science engineering. Although we did not specify a programming language, all students chose to work in Python, the standard language for the course. All essays were collected, and all but one pair (15 of the 17) students also consented to be interviewed shortly after completing their essays. This resulted in a total of about 10 hours of interview data. Interviews were semi-structured, with prompts that asked students to walk the interviewer through the development process of their essay, reflect on the connection between computation and learning physics, and reflect on the ways in which computation enabled creative exploration in the courses they were taking. The interviewer posed initial prompts, clarification questions, and follow-up questions, but interviewees did the majority of the talking. Because students were native Norwegian-speakers, interviewees were given to option to speak English or Norwegian depending on their preference and three groups chose to conduct the interview partially or entirely in Norwegian.

We then engaged in thematic analysis [47] of the interviews and coded them for themes related to the three pillars of computational literacy. We also analyzed the computational essays, looking for ways in which they supported the self-reported data from the interviews, as well as places where the students addressed issues or ideas not discussed, like specific coding practices, writing styles, and report structures. During the initial round of analysis, the first author used a process of analytic memoing [48] to note emerging themes and patterns in the data, which were then interpreted in terms of the three pillars of computational literacy. Next, the first author coded the entire dataset in NVivo using these themes as a starting point, after which he refined the code categories by renaming, collapsing, and/or splitting categories with a large or small number of entries. The second author then reviewed and coded a subset of the data (two paired computational essays and interviews that the first author judged to be rich examples of computational literacy), and the two codings were compared, with discrepancies being resolved through discussion and/or modification of the framework. Finally, the first and second author chose a series of excerpts from across the data corpus to illustrate the final theoretical categories. These excerpts are presented in our results section below.

We chose thematic analysis as a method in part because it matched our research goals—we are interested in capturing and distilling a snapshot of computational literacy in situ, in order to flesh out the theory of computational literacy originally proposed by diSessa. As mentioned, we approached this analysis through the lens of an existing theoretical framework. However, in our analysis, we also looked for opportunities to identify themes would let us expand and apply this extant framework to the specific domain of computational physics—that is, we wanted to be open to what we might find in the data. We found that thematic analysis allowed for this blend of using an existing framework while being willing to consider new themes that might emerge. Again, we emphasize that our goal is exploratory, as befits an initial case study, and for this reason we illustrate our framework with quotes and excerpts from the data for each sub-category but do not count frequencies of each category in our dataset or measure inter-rater reliability.

## IV. RESULTS: A FRAMEWORK FOR COMPUTATIONAL LITERACY IN UNDERGRADUATE PHYSICS

In the course of this analysis, we have identified a variety of different elements that are associated with the material, cognitive, and social computational literacy of undergraduate physics majors. However, similar to other previous work on computational thinking [49], we see these elements as existing at different theoretical levels, ranging from concrete practices, to necessary bodies of knowledge or ways of thinking, to general beliefs about the purposes of computation in physics. So, we have grouped them together and organized them in order of increasing complexity (and abstractness), from "practices" to "knowledge" to "beliefs." Our final organization is shown in Figure 3. Before elaborating the framework, we clarify what we mean by the categories of practices, knowledge, and beliefs to facilitate a clear introduction to components of each pillar.

By *practices* we mean the observable activities in which a student engages—essentially, the actions they take, which reflect respective aspects of their physics computational literacy. Practices may be found at a variety of grain-sizes, from specific, concrete actions like writing loops or creating graphs, all the way up to general patterns of behavior like tinkering, debugging, or modeling. Analytically, we inferred these practices from student artifacts (their computational essays) and descriptions they provided in interviews of specific actions or approaches they took to coding, analysis, and communication.

By *knowledge* we mean a students' thinking about particular concepts or ideas. Again, in this framework for physics computational literacy, such knowledge and thinking could occur at a variety of grain sizes, from facts or concepts to cognitive strategies and higher order ways of thinking. Speaking of knowledge in this way suggests an epistemological assumption on our part, that students can in some sense obtain and possess certain bits of knowledge. This cognitivist assumption is consistent with the overarching theory of computational literacy as described by diSessa [5]. We acknowledge potential limitations of this perspective but maintain that it is useful to be able to speak of certain information or concepts that a student might know or understand. Analytically, we inferred student knowledge from cases in which students justified or generalized their practices, either in interviews or in textual explanations. In other words, we did not assume that students were always cognizant of the reasons why they were using certain practices, or the meanings and assumptions underlying them, until they explicitly referenced this knowledge in essays or interviews. However, evidence of practices and evidence of knowledge frequently went together, because students would often write a block of code in their essays, then explain the reasoning behind that block of code in the next line.

By *beliefs*, we specifically mean student beliefs relating to computation, both generally and at the intersection of the domains of computation and physics—for example, the epistemological status of knowledge gained via computation, the uses of computational tools, and preferences for particular computational languages and applications. Thus, we characterize something as a belief if it is associated with an attitude or a value judgment about computation in some way. We inferred these beliefs based mostly on evidence from the interviews, which was then supported by examination of the essays. For analysis, we categorized statements as beliefs when they addressed students' large-scale views of the nature of programming, computational physics, and the ways in which one should communicate with and about code.

We have presented these three pillars as separate to facilitate discussion of each pillar. However, as we have noted, these pillars are interrelated, so in practice there were seldom elements of data that fell exclusively into one pillar or another. Because of the design of the project, elements from the three pillars nearly always overlapped—that is, whenever students wrote code (a material practice), it was in the service of a physics modeling project (cognitive), in the context of a communicative document (social). In interviews, students would verbally communicate (social) about the code they had written (material) and the conclusions they had drawn from their project (cognitive). We would, however, characterize this overlap as a feature of the research design, rather than a bug—we designed the computational essay project to provide a rich snapshot of students' computational literacy in practice, and so it was unsurprising that most of the evidence we collected showed hallmarks of two or even all three of the pillars. In our illustration of the framework below, we have tried to decouple these three pillars as best we can in order to foreground examples that show features unique to one pillar or another.

Based on these theoretical distinctions, our final categories are shown in Figure 3.

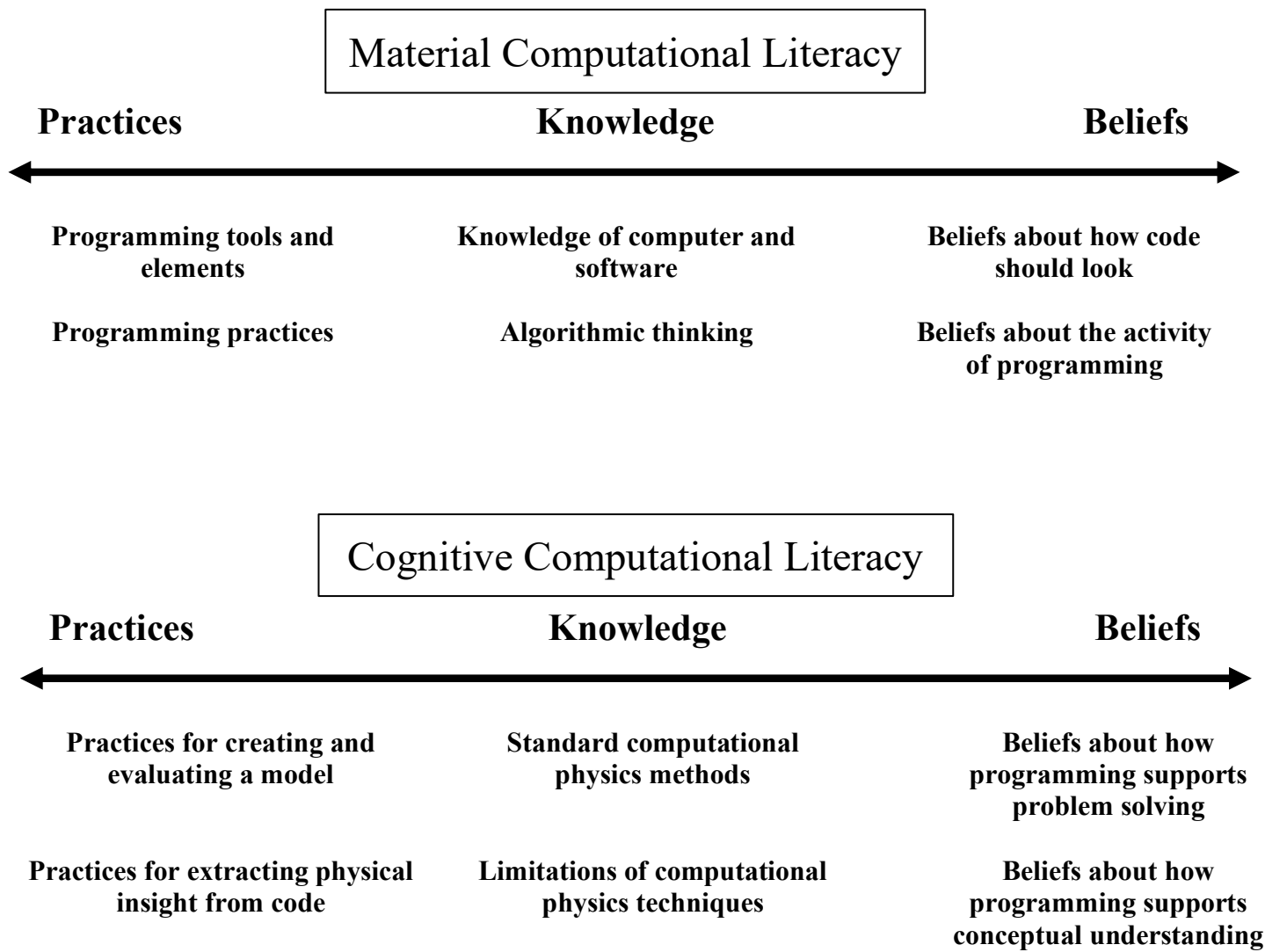

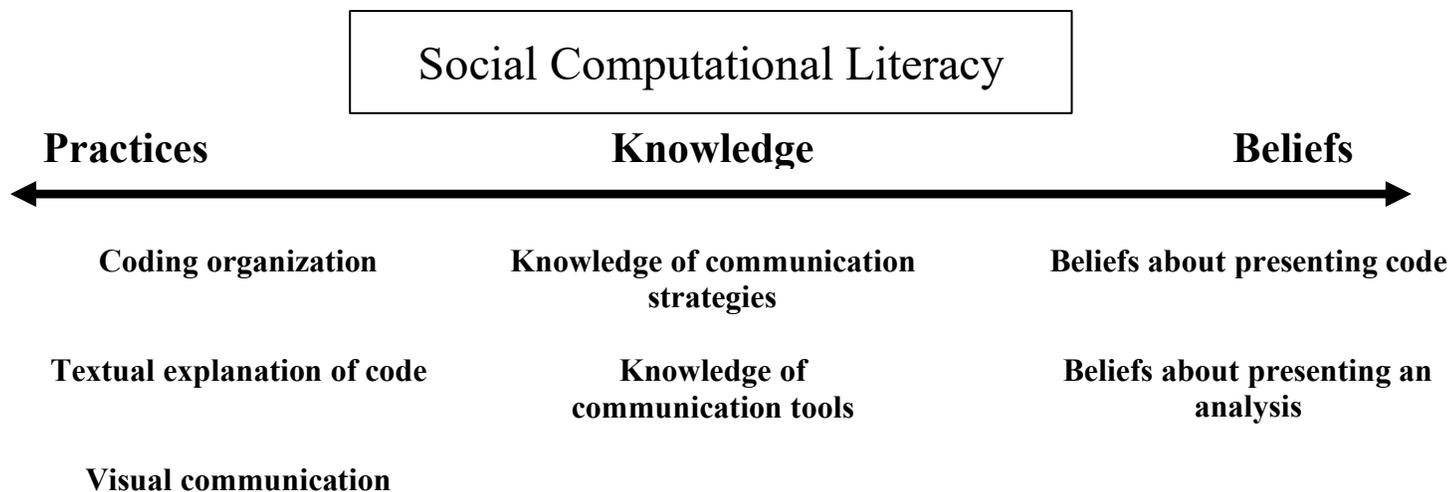

Figure 3: Identified elements of material, cognitive, and social computational literacy in an undergraduate physics context. Elements are organized on a continuum ranging from concrete practices, to knowledge elements, to higher-level beliefs.

In what follows, we elaborate on these different identified aspects of computational literacy and exemplify them with excerpts from the collected interviews and computational essays. Our goal with these examples is to illustrate each category, rather than provide a thorough accounting of all of the ways each could (or did) manifest in our data. In cases where excerpts have been translated into English, the translation has been noted after the text.

### A. Material Computational Literacy: Core Programming Skills

The category of material computational literacy encompasses the practices, knowledge, and beliefs necessary to be able to work with computation and code. Although these material elements are likely to be used by students in a variety of disciplines, we wish to highlight this subset of practices, knowledge, and beliefs as especially relevant to undergraduate physics.

#### 1. Practices: Programming elements and practices

The practices aspect of material computational literacy concerns the mechanics of coding—the skills and procedures necessary to read and write code, regardless of language or discipline. Although some might consider them to be basic, these practices form the foundation on which the rest of computational literacy is built. Just as one cannot be considered print literate if one cannot read, write, and construct sentences, one would not be considered computationally literate without some amount of material programming skill.

In our dataset, *basic coding elements* included familiarity with the building blocks of code: loops, variables, functions, packages, toolsets, and syntax. Within the collected computational essays, all examples utilized some combination of variables, loops, functions, and Python packages (such as the plotting library matplotlib). Students also demonstrated competence with basic data handling, including the creation and use of arrays to store and manipulate numerical results. Students additionally showed familiarity with several different developer environments, including Jupyter notebooks (used for their computational essays), and self-reported use of two other environments, Atom and Spyder.

*Programming practices* encompasses the basic activities one does while coding, regardless of discipline or context: for example, troubleshooting and debugging code, tinkering with code, data handling and storage, and building on others' code. Note that we are theoretically narrowing our focus, here, to cases in which students are primarily trying to understand and work with the behavior of code rather than the behavior of a simulation. Although we were not able to observe these practices directly while students were writing their essays, students regularly commented on them during interviews. For example, several students brought up their practices of tinkering and debugging, especially when they were trying to understand the behavior of the example code they had been given. In one interview, Jeffry said the following:

> **Jeffry:** I think the most difficult part was getting it to work, basically. Debugging, and trying to figure out why it didn't work. That was probably the most difficult part because there's a... The problem arises when the error is not in the math or in the reasoning but in the numerical implementation or in... When it's a numerical error. Because everything looks like it should work but it doesn't. So trying to and test each element separately to figure out what works and what doesn't work.

Here, Jeffry explicitly distinguishes between errors in the math and reasoning (which we would characterize as cognitive) and "numerical error," which he seemed to use to refer to simulation errors caused by his particular code. Jeffry's process of "trying to test each element separately," we argue, is therefore essential to the material aspect of physics computational literacy, because code is seldom completely correct as written the first time and often requires some amount of adjustment to run without bugs or errors. In order to disentangle unexpected behavior due to coding errors from that arising out of physics errors, computationally-literate physics students must develop a certain degree of material competence in programming practices like debugging.

## 2. Concepts: Algorithmic thinking and knowledge of the computer

These basic programming practices require a certain amount of higher-level conceptual knowledge to support them. Within our dataset, we classified these concepts into two categories: knowledge of higher-level computational techniques and algorithms, and knowledge of the underlying behavior of the computer (sometimes known as the "notional machine" [50]).

We use the term *algorithmic thinking* to describe students' understanding of how to combine the various basic tools of programming to iteratively accomplish specific tasks like search algorithms or optimization procedures. We realize that the term algorithmic thinking comes with many theoretical connotations stemming from the various definitions of algorithmic thinking in the computer science and mathematics education research literature (e.g., [51–53]). However, here we simply take it as a way to discuss students' thinking about coordinating programming tools and algorithms during the coding process. For example, one pair of students in our study (Frank and Ian) used an iterative search procedure in order to calculate the necessary amount of current required to accelerate a train up to a given velocity using a railgun mechanism. In their interview, the pair described their procedure as follows:

> **Frank:** These two loops are for the case where the acceleration is bigger than the desired acceleration, and the velocity is smaller than the desired velocity, in which case we adjust down the current. And if the acceleration is too small we increase the current. And the last loop is if… if it reaches the desired speed, then it calculates how much current is needed to keep the net force equal to zero.
>
> **Ian:** And the reason we have this… if the acceleration is too large or too small, and the speed is smaller than what we want… Because to begin with we tried it without the loop that checked the speed, but then we saw that when it reaches the desired speed it always tries to decrease the acceleration. And then in the next iteration it tried to increase it again, then decrease it, and went back and forth the entire time. (translated)

This type of algorithmic thinking is essential to physics computational literacy because many computational techniques are algorithmic in nature. Improvements or modifications to the techniques often involve minor modifications to the algorithms, making it essential that students are able to follow the step-wise logic of such techniques. This example also demonstrates the entangled nature of these three pillars. We are focusing on the material pillar in this example, because the students' discussion of the iterative nature of loops evokes a key component of their material computational literacy. However, we note that they are also talking about these loops in relation to physics concepts (velocity, acceleration, and current). So, although we are foregrounding the material aspects of this this example, we do not claim that they are engaging exclusively with the material pillar.

*Knowledge of the computer* involved an understanding of the computer's behavior and how it could affect the execution of written code. This knowledge can essentially be thought of as a student's mental model of how the computer is behaving, including the relevant aspects of the machine (e.g., memory usage, processor speed, RAM) and the process by which it executes code (e.g., compilation and stepwise execution of instructions). This aspect of material computation literacy differs from basic programming elements in that knowledge of the computer has to do more with the fundamental hardware and software elements that determine how the computer actually works. In the computer science education literature this kind of mental model is often referred to as the "notional machine" [50].

One primary way students demonstrated their understanding of the computer was in their discussions of data structures and how they could limit or affect the behavior of simulations. For example, several students encountered difficulties due to their simulations running slowly or crashing after a certain number of iterations. Often these difficulties stemmed from the computer's data-handling procedures, requiring students to try to understand those procedures in order to diagnose the problems with their simulations. Students addressed these issues in different ways; for instance, one student, Mel, greatly increased the speed of his simulation after he put all of the values in his dataset into a series of arrays. He then used those arrays to carry out calculations that had previously been done individually in nested loops. By cutting down the number of operations the computer had to do, he greatly increased the speed and efficiency his simulation. A snapshot of his code, in which he uses this procedure to solve Poisson's equation with an approach called the method of relaxation, is shown in Figure 4 below:

```
In [119]:

def poisson(potarray,ds,rho,Vybound):
    #Vybound should be a nx2 matrix where j indexes 0 and 1 are respectively bottom
    eps0=8.854e-12            #the vacuum permittivity

    V=np.copy(potarray)
    lx,ly=np.shape(V)
    Vnew=np.zeros((lx,ly))    #initializing an array with new potentials
    rhof=rho*ds**2/eps0       #the term containing rho in the equation above

    #Calculate the 1/4 factor shown above using periodic boundary conditions on the
    Vnew+=0.25*(np.roll(V,-1,axis=0)+np.roll(V,1,axis=0))    #now Vi+=0.25*(V_i+1 +
    Vnew[:,1:-1]+=0.25*(V[:,0:-2]+V[:,2:])                   #now Vj+=0.25*(V_j+1 +
    "Solving the y-boundaries separately:"
    Vnew[:,0]+=0.25*(Vybound[:,0]+V[:,1])
    Vnew[:,-1]+=0.25*(Vybound[:,1]+V[:,-2])
    #Add the rho term:
    Vnew+=rhof
    return Vnew
```

**Figure 4:** Example of array-based calculations, implemented to solve Poisson's equation through the method of relaxation

This snapshot, along with Mel's justification of it in his interview, suggests that he was well aware of how different code structures would lead to different numbers of steps in an algorithm, which would in turn affect the time taken to run a program. Such an understanding of computer behavior is a necessary component of physics computational literacy, especially in higher level applications like many-body simulations, where computing time quickly becomes a limited resource.

### 3. Beliefs: Beliefs about programming as an activity and how code should look

In addition to these practices and knowledge, the students also displayed some more tacit beliefs about the purpose and uses of programming. These fell into two categories: beliefs about the activity of programming, and general beliefs about how code should appear (characteristics of "good" code). Under the category of *beliefs about the activity of programming*, students frequently expressed preferences for different programming languages, based on their perceived uses and affordances. For example, Casey had the following to say about Python vs. MATLAB:

**Casey:** Python is more versatile, kind of. I mean I'm sure MATLAB has its really good strengths, but it's just for math, so I like that Python has a combination of making good programs and also doing calculations so we can personalize a bit more Statements like these reveal that students see programming as generally useful in different ways—for example, viewing programming as primarily a tool for performing calculations versus writing simulations.

Students also discussed preferences about how they should approach the activity of programming. For example, one choice students faced was the degree to which they would rely on the pre-written code they had been given versus writing (or re-writing) code themselves. While many students chose to use the code they'd been given, Mel explicitly expressed a preference for re-writing code in order to give himself more opportunities for creative freedom:

> **Mel:** They put out the example codes that you can build on, but I decided to do the code myself. So that's a lot of coding, and a lot of creativity choices. So in that way, it's similar to this computational essay project, because you can decide on how to solve the problem.

Students also hinted at their *beliefs about the how code should appear and be structured*, both in interviews and through the code they wrote for their computational essays. In general, students seemed aware of commonly-taught "good coding practices" [54] such as defining functions to avoid unnecessarily copying code, and commenting code to make it transparent to readers. These beliefs likely stem from the training students received in coding during their first semester; for example, Arthur described his training in programming as imparting the following values:

> **Arthur:** How to use for and while loops effectively, how to not repeat yourself in the code and how to make sure that when you look at your code two months later that you actually know what's going on.

These values seemed to be reflected in the students' computational essays, though it should be noted that the essays were explicitly meant to encourage students to document both their thought processes during the project and the code they had written.

### B.  Cognitive Computational Literacy: Using programming to solve problems and understand the world

Because we are interested in developing theory around computational literacy in physics, the category of cognitive computational literacy focuses on the particular domain of physics—it encompasses the ways in which physics students use programming to solve problems and extend their understanding of the world. In order to distinguish these cognitive elements of computational literacy, we looked for computation-related practices, knowledge, and beliefs that students might use to further their physics understanding, regardless of the specific programming implementation (material elements) they might use.

#### 1.  *Practices: Practices for modeling and extracting physical insight from code*

Although there has not yet been a systematic study of the programming practices used in undergraduate physics, some initial work has been done in this area by the AAPT Undergraduate Curriculum Task Force [3]. In our dataset, we observed examples of all the coding practices documented in that report, however we will not unpack all of them here. Instead, we are grouping them together into two categories: modeling practices, and more general practices for extracting physics insight from code.

Due to the specific set of topics and types of research questions common to physics, many of the practices common to the discipline are focused on creating, refining, and evaluating computational models. Students displayed a variety of these *modeling practices*, including various ways to define and estimate model parameters, make simplifying assumptions, and check the reasonability of their results. For example, one pair of students, after creating a simulation of a train powered by a railgun mechanism, re-used that code to simulate the launch of a cannonball as a way to verify that their code was producing reasonable results. Students also reported frequently consulting external sources, both to compare their results to published values and for inspiration on model development.

A number of students also described how they would often tweak variables and "play around" with systems to understand the effects of different physical parameters. For example, Morton contrasted computational and analytical physics assignments, arguing that computational simulations presented more opportunities for this kind of tinkering:

> **Morton:** You can play around, a lot more. And I guess if you, if it isn't working like you want to you can always, mmm—how shall we say? Fiddle around and try to just add some mysterious factors here and there to see if you can locate a problem in the way you're thinking or, yeah. So, kind of just fiddle around, which you can't really do in an analytical situation because there's wrong and there's right.

Because Morton is discussing how he tweaks simulations to improve his understanding of the system ("locate a problem in the way you're thinking") rather than find errors in the code, we consider this practice to be primarily cognitive in nature. This process of "fiddling around" with variables is a key part of the model refinement process [14], and has also been previously argued to be a general hallmark of computational literacy [35,55].

At a broader level, there are certain general *practices for extracting physical insight from code* that the students relied on when performing their analyses. We have chosen to distinguish these from modeling practices because they were focused less on the construction and use of models, and more on the ways in which students leveraged the particular affordances of code to build their understandings. One such practice includes the creation of visualizations of abstract phenomena, either physical (e.g., magnetic fields) or mathematical (e.g., functional behavior). All students in the present study included at least one code-produced visual in their computational essay, most frequently plots of particle trajectories and other associated quantities (like energy) as a function of time. However, students also described how this visualization could be an intermediate step in their investigations:

> **Jeffry:** Often when I do projects with computation, I have to make many plots that are just intermediary. Plots that are not really part of the results but just to get an idea of how far I've come. Look and see. If I've found an expression, I can plot it and say, "Does this look realistic or is it completely wrong?" That's not always easy to see just by looking at it.

Beyond visualization, another computational practice common to undergraduate physics is "repeating a computation many times using different sets of parameter values of particular interest" ( [3] p.7). Several groups demonstrated this practice, leveraging the iterative nature of computation by repeatedly running simulations to gain insight on their system and explore the effects of random variation on simulation results. For example, two of the groups who wrote their computational essays on the topic of lightning both reported repeatedly running their simulations to gain an understanding of the typical locations of lightning strikes under varying conditions. Although this practice is similar in nature to the algorithmic thinking practice described above, we associate it with cognitive computational literacy because it is explicitly focused on helping students understand the physics of a simulation rather than the code.

All of these practices, we argue, represent ways in which computation can concretely contribute to students' understandings of physics. Because visualization and plotting are fairly simple to do computationally, they represent a major affordance of using computation for doing physics. Iteration is a relatively simple computational practice, which (if used properly) can yield great insights into physical systems. The process of building, refining, and evaluating models is essential both in learning physics and in the discipline as a whole [56], and practices like those described above are central to the modeling process. However, to leverage these computational affordances, students must learn how and when to engage in these practices.

### 2. Knowledge: Computational physics techniques and their limitations

In order to use programming effectively, students also must have certain bodies of knowledge related to the ways in which computation is used in physics. This category overlaps with the knowledge category of material computational literacy, in that it is focused on the applications and limitations of computational algorithms and techniques. However, this body of knowledge specifically focuses on the meanings and applications of these techniques to analyzing physical systems.

The first category of knowledge we have identified is *standard computational physics methods.* At the University of Oslo, students are taught a variety of computational methods for simulating equations of motion, including integration algorithms such as Euler, Euler-Cromer, and Runge-Kutta. Most of the students expressed familiarity and comfort with Euler-Cromer integration, having used it frequently in their mechanics course the previous semester. A few also named higher-order methods, such as Leapfrog integration, and two groups actually implemented Leapfrog integration in their essays (both of which were variations on the magnetic trap template we had given them).

As the complexity of physics problems one is trying to solve increases, it is not enough to just know an algorithm—one must also know its domain of applicability, which requires *knowledge of a technique's limitations*. That is, each algorithm has certain affordances and certain drawbacks: Euler integration, for example, has the advantage of being computationally cheap (it requires relatively few computational steps) but does not conserve energy in simulations, causing unphysical behaviors over long simulation times. Higher-order integration algorithms may do a better job with conserved quantities, but require more computational steps. For this reason, the choice of algorithm has significant consequences for the behavior of a simulation, and one primary area of cognitive computational literacy in physics is a knowledge of these limitations. Several students explicitly mentioned this topic in interviews. For example

Morten and Kurtis had the following exchange in discussing their essay, which simulated an attempt to trap a high-energy particle with a magnetic bottle.

> **Morten:** We—we did use the Leapfrog [integration algorithm] […] But it's was like, twice as slow, almost.
> **Kurtis:** And it [the simulated particle] seemed to, then, spiral inwards. I think that's what it did? I'm not sure.
> **Morten:** But that, that could be a physical thing, though.
> **Kurtis:** Because the energy isn't actually conserved. Because you get more kinetic energy. So it's all just, yeah. But yeah. We didn't try it in the end when we had the finished result, so then we could have compared the different methods. Euler-Cromer, that seemed to not work too because it really has to be at a tangent. I think that's why that one didn't work.
> **Morten:** I guess maybe we should have used something even higher order? I don't know.

Here, we see Morten and Kurtis explicitly contrasting the standard Euler-Cromer method (which they had frequently used in their courses) with the more accurate Leapfrog method that they had used in their essay. The pair seemed well aware that different integration methods could have varying effects on simulation results, specifically naming energy conservation as a limitation, and they even suggested comparing multiple integration methods as an extension of their analysis. We would argue that this level of understanding, which goes beyond just the procedural ability to implement certain standard computational physics techniques, is an important component of students' physics computational literacy.

### 3. Beliefs: Beliefs about how and when one can use computation to improve one's understanding of physics and the world

Under the category of *beliefs,* we identified a set of beliefs held by students about how they could use computation to improve their understanding of physics. These fell into two categories: beliefs about the uses of computation for problem solving, and beliefs about the uses of computation for building conceptual understanding.

Under the category of *beliefs about the uses of computation for problem-solving*, several students described a belief that computation empowered them to solve more difficult problems and model more complex situations. For example, Lee said addressed this topic in the following way:

> **Lee:** When you have the opportunity to model something [with computation], then you can make the situation a lot more complex than if you were just going through it on paper or by hand or in your head.

Here, Lee argues that computation allows one to address situations (problems) that are significantly more complex than those that are approachable analytically or conceptually. A large part of this complexity lies in the fact that certain factors, which make analytical problems intractable, can be easily included in computational simulations. Jeffry explicitly commented on this aspect of computation:

**Jeffry:** In [some] situations there aren't really any good analytical method because the analytical—finding an analytical solution will require making lots of simplifications. [...] If I want to find a more realistic solution, it's often much easier to write a program than to try and figure out how to incorporate everything into an analytical solution.

We find Jeffry's statement to be particularly interesting because he argues that computation can actually simplify the problem-solving process when compared to analytical approaches.

In addition to this focus on problem-solving, students also expressed certain *beliefs about the uses of computation for building conceptual understanding.* For example, Lily described the way computation allowed her to visualize abstract concepts:

**Lily:** I think that's because it's something you don't physically see, so it's just a concept. Then whenever we were able to take a system ... I can't remember what it was, but plot the kinetic energy and then see the motion and potential energy, then you could actually make the connections and it was visible to see so that when I was thinking about it later I could kind of just imagine that in my head again.

Here, Lily argues that the different representations produced by a program allowed her to draw enduring cognitive connections between the energy of a system and its motion.

Together, these excerpts show evidence of a set of larger-scale beliefs that students hold about the usefulness of computation for their physics learning. These beliefs are important in part because they are likely to influence both the degree of student "buy-in" to computation in physics courses and the ways in which they might use computation during their studies and beyond. It seems fair to assume that students who feel that computation is especially useful for problem-solving will be more likely to actually use it when given a choice as to how to solve a physics problem. Similarly, students who feel that computation is a viable avenue for developing their intuition about a system may be more inclined to use the visualization and modeling practices previously described.

### C. Social Computational Literacy: Communicating with and about computation in physics

Social computational literacy covers the ways in which students communicate with and about computation. Both physics and programming are social endeavors with their own norms and practices for communication. So, the set of practices, knowledge, and beliefs within this pillar of computational literacy focus on the ways in which students communicate with others about the structure, purpose, and results of their code, specifically within the domain of physics.

#### *1. Practices: Organization and explanation of code through text and visuals*

We have identified several social computational practices that students use when communicating about their programming. These include code organization, textual explanations, and communication through visualizations.

*Code organization* encompasses the ways in which students try to make their code clearer to a reader. For our students, this included "chunking" code into groups such as import statements, initial conditions, model parameters, and integration loops. Most students also defined functions (or use pre-defined functions) for key, physically significant aspects of their models (e.g., magnetic forces) in order to cut down on the amount of code in any particular block and avoid excessive copy-pasting.

*Textual explanations* include both in-line comments within blocks of code and written explanations outside code blocks to describe the purpose and output of the code. Although all students commented their code, the textual descriptions ranged in quality and detail, from sparse or absent in some notebooks to thorough documentation in others. We show one example of a thoroughly-documented block of code (including textual explanations, and two types of in-line comments) in Figure 5.

We import the standard libraries and define the input variables for the potential to calculate the initial voltage.

```
In [1]:
import numpy as np
import matplotlib.pyplot as plt

"""Defining variables for LINAC2"""
c = 2.99*10**8 #Set the speed of light
q = 1.6e-19 #Set the charge of the particle to the charge of a proton
m0 = 1.67e-27 #Set mass of the particle to the mass of a proton

v0 = 0.3*c #Set wanted exit speed of LINAC2
l0 = 500 #Set length of LINAC2
U0 = 1/2*(m0*v0**2)/(q) #Set voltage of LINAC2
print("Voltage of LINAC2 is: %e V" % U0)

Voltage of LINAC2 is: 4.199053e+07 V
```

**Figure 5:** Example of thoroughly-documented code, including textual explanations, general inline comments (red) and specific comments (blue).

Finally, we noted that in addition to using visualizations to bolster their own understanding (as described in the cognitive practices section above), all students also *communicated through visualizations* in their essays to illustrate certain computational results and ideas. Many of these visualizations were plots generated by the code itself, ranging from straightforward plots of parameters like trajectory or electric current with time to heat-map plots of electric potential and magnetic field strength. However, students also frequently included pictures or diagrams in their computational essays to illustrate the systems being modeled. In interviews, some students additionally described the ways in which they tweaked or adapted their figures to make this communication more effective; for example, Arthur, who wrote his computational essay on a magnetic trap called a Biconic Cusp, described how he had modified his plot to more effectively show variations in the magnetic field strength:

**Arthur:** [I] changed the title of the figures. That was an important one. Added a color scheme to the strength of the magnetic field by, I think, just using a norm function on the vectors, so I got the vector length. And plotting it as a scalar field with a color bar.

Although simple, practices like these are an important part of physics computational literacy, because they allow students to use their computational simulations for more than just building their own personal understandings. They give students opportunities to communicate their methods and results with other students (facilitating collaboration) and with teachers/supervisors (facilitating evaluation and feedback). They also reinforce good habits from the material and cognitive pillars, like structuring code in an efficient way and facilitating interpretation of physics results. However, practices like these do not arise naturally. Just as students need to be trained in clear and effective writing, they must also be trained in effective code organization and explanation. We argue that computational essays offer an ideal environment to both train and evaluate students on these types of skills, because they are designed to encourage students to share both the methods and results of their simulations

## 2. *Knowledge: Knowledge of communication tools and strategies*

The category of social computational knowledge encompasses the knowledge required to communicate with and about computation. Within this category we include knowledge of specific communication tools and their operation, as well as knowledge of effective communication strategies. While these categories will theoretically be situated within students' more general modes of communication, here we are considering the ways in which students communicate specifically about physics and computation.

By the time they reach college, most students have at least a basic familiarity with some kind of word processing software, and all students will have some experience with communicating through handwritten text and equations. However, the inclusion of computation in a physics curriculum adds another level of complexity to the art of communicating because computer code is meant to be both read and executed. In order to not just execute code but also communicate with and about it, students need facility with tools that allow them to mix both code and text, such as the Jupyter notebooks used in this study. However, in interviews most students reported little familiarity with notebook software prior to the project. For this reason, we have identified *knowledge about communication tools and their operation* as one important area of social computational literacy knowledge. As an example, Perry described his developing understanding of Jupyter notebooks as follows:

> **Perry:** I think the idea is cool, that you're mixing together text, code, pictures, right? And it comes through… you get a nice mix, then, right? And it's interactive. I think it's a bit tricky to get started with. If one doesn't know all of the keyboard shortcuts then it takes a while to figure out exactly how you make things happen. You have to look some things up. So that's a bit tricky. It's the first time I've used it. And also, the order you execute the program, right? And suddenly, if you execute something you shouldn't have, and you overwrite some variables, right? […] I thought it was nice to try, at any rate. (translated)

Here, Perry explicitly comments on the different affordances of computational notebooks—they allow one to mix different mediums of communication (code, text, pictures)—but also describes the challenges associated with the tool, including the non-linear nature of notebooks. It is this familiarity with both these affordances and potential difficulties with the tool that constitute this aspect of social computational literacy.

Students reported varying levels of notebook use. Although all participating students turned in computational essays written in notebooks, some students used them more often than others during the actual development process. That is, some students reported writing all of their code in a separate environment then copy-pasting it into the notebook, while others said they iteratively developed their projects in the notebook itself. A few students even explicitly used the non-linear nature of the notebook in their communication: one student, for example, capitalized on the interactive nature of the notebook medium and flagged locations where users could modify the parameters in his simulation of lightning and re-execute the cell to witness interesting results.

Beyond familiarity with communication tools, in order to be successful in the project students also had to have some *knowledge of communication strategies*. In part, this included the ways in which students structured their essays. Most students drew on the standard structure of a scientific report to some degree, with a defined introduction and conclusion. Because the students had been given an example essay as a reference (see Figure 2), it is not surprising that many of them followed the overall form and structure of this example. However, some students took this one report structure step further, with one student even thanking the teaching team in an "acknowledgments" section.

Another aspect of these communication strategies concerned students' attention to the "audience" for their projects, both the readers of the essay and the literal audience to their presentations. This mostly affected the degree to which the students explained their code and the development of their model; some students went into great detail on their methods, while others had sparsely-documented code sandwiched between an introduction and conclusion. In interviews, some students commented on these different norms; for example, Frank shared the following reflection on the different types of presentations he witnessed:

> **Frank:** As I'd originally understood it, before the other presentations, there was going to be more of a focus on how the code worked. And then we saw some of the other presentations, and there they just skipped by… It was like, they scrolled by the code, and just showed the plots, discussed the plots, and then scrolled by the next… They explained, in a way, how the code worked, and said "here we adjusted the magnetic field to be stronger," for example, and then they scrolled by the code and showed the plot. So, I think if the idea behind this is to explain the code, it could… it would be good to use the Jupyter notebook, where one can change small pieces of the code in one's preferred order. (translated)

Here, Frank seems to be commenting on the fact that certain students focused mostly on the output of their code, while he had interpreted the project as explicitly asking the students to provide detailed explanations of their code. Although we, as project designers, preferred greater detail in students' code documentation, we feel that the students' degree of documentation also speaks to their attention to the different audiences for their projects—some students explained their code and modeling process in great detail, aiming it at a more general audience, while other students seemed to assume an audience experienced in computation and physics.

We see this type of social computational knowledge as an important aspect of physics computational literacy because it affects the ways in which students choose to communicate about their code and investigations. Even if students have mastered the practices mentioned above—code organization, textual explanation, and visualization—it is no guarantee that they will be able to produce a coherently-written essay or narrative. Therefore, it is important to help students develop the necessary skills and knowledge to be able to communicate in these disciplinarily-authentic ways.

### 3. Beliefs: Beliefs about presentation of code and analyses

Our final category under the social pillar of computational literacy encompasses students' beliefs about when and how they should communicate about their code and projects. Within this category, we identified two sub-categories: beliefs about presenting code, and beliefs about presenting computational analyses.

Under the category of *beliefs about presenting code,* we found that students' written work sometimes revealed tacitly-held beliefs about the reasons and ways in which one should document one's code. In part, this involved their use of the Jupyter notebook tool, with some students treating the tool as a log of their inquiry process and others using it to produce polished, shareable reports. Those who used notebooks as logs tended to use less textual explanations, and the text that was included was mostly used to record their procedures and conclusions. Their code tended to be messier, and some students even left intact dead-ends and segments of broken code. In contrast, students who used the notebooks to produce polished reports focused their text more on the theoretical underpinnings behind their work, as well as justifications for their procedures. Although one of our goals with the design of the computational essay project was to encourage students to produce polished reports, it is worth noting that professional scientists commonly also use notebooks for both shareable reports and exploratory analyses [45]. So, we feel that both of these uses are authentic to the discipline of computational physics.

Some students also explicitly discussed their beliefs about code documentation in their interviews, especially with regard to familiarizing themselves with the example code they had been given. For example, Rupert shared the following critique of the example code:

> **Rupert:** For my part, I would have preferred some small start-comments on what the function was actually doing, and what the function returned, which I felt were kind of missing. Maybe a little more… a few more comments about what it was doing. There were some small in-line comments, but yeah. (translated)

Here, Rupert is arguing that the small in-line comments included in the example code could have been augmented with additional documentation on the specific purpose of defined functions. Statements like these reveal that at least some students have beliefs about the nature and practice of code documentation, which in turn are likely to influence the procedures and knowledge that they use.

Under the category of *beliefs about presenting an analysis*, we found that students expressed a variety of beliefs and framings towards the process of communicating their project results. For example, several students either tacitly or explicitly showed signs of viewing the project as essentially a mini research-study, including the steps of defining a research question, reviewing relevant literature, creating and executing a methodology, analyzing results, and writing up conclusions. One strong piece of evidence for this framing was that in interviews, certain students discussed the difficulty of finding a good "research question." For instance, Casey described his group's development process as follows:

**Casey:** We got an initial code and essay with just a plot of how the magnetic lines look like and how a particle with a certain initial velocity will move inside the bottle, or not move, or get trapped, obviously. So we had to come up with a research question where we could explore this.

Students also described engaging with the relevant literature (both scientific papers and sources like Wikipedia), and most included at least one reference to an external source in their essays. As mentioned above, students also brought this framing to the structure of their written essays, often structuring them in a similar fashion to scientific papers, including sections for references and acknowledgments.

Although many students drew on this research-project framing, some students incorporated an element of more personal exploration, sometimes including reflections on difficulties they encountered or discoveries they had made along the way. For example, Rupert and Iris included the following reflection in their essay, which modeled the effects of conductive lightning rods on lighting strikes:

> *We have to calculate the surface potential of the conductor first, this might seem daunting at first, we did a lot of back and forth to figure this out to be honest, until we realized that we only need to deal with a constant potential.*

We argue that it is important to be attentive to these types of beliefs because they are likely to affect which practices and knowledge the students use. That is, if students do not have productive beliefs about the utility of code documentation, they are unlikely to spend much time documenting their code. Similarly, without some kind of beliefs or framings for how to present a computational analysis, students are less likely to draw on their learned practices or knowledge for communicating computational results.

## V. DISCUSSION

Our goal in this study has been to present a framework to help understand how computational literacy can manifest in undergraduate physics. This is by necessity a limited framework, as it is only focused on undergraduate students and so does not encompass the diversity of practices, knowledge, and beliefs of professional scientists. Additionally, the data that informs the framework was based on a small sample of students, and so is by its nature exploratory. Stated simply, we mean this framework to be a first attempt at exploring the space of computational literacy within physics, to be elaborated and fleshed out in future studies.

We also fully acknowledge that this framework is strongly tied to the context in which these students are learning computation, which will greatly affect the types of practices, knowledge, and beliefs they demonstrate. In part, this is an expected component of the theory, because the ways in which literacies manifest are always strongly tied to the cultural and historical contexts in which they are used [20]. Computation and programming are also not yet widely integrated into physics education [15,16], and so the ways in which computational literacy will manifest in other programs is likely to be different than the ways it manifests at the University of Oslo. At the same time, we feel that the University of Oslo can present a useful example of a program in which computation has been integrated for close to two decades, and the potential benefits and opportunities this long-running integration brings. For this reason, we argue that this project functions as a useful case-study of the ways in which computational literacy *can* manifest in practice—essentially, an existence proof for the theoretical construct of undergraduate physics computational literacy.

We see three primary uses for this framework. First, theoretically, we wish to use this framework to argue for the inclusion of (and attention to) students' computational practices, knowledge, and beliefs across the three pillars of computational literacy for future studies of computational physics education. Second, more practically, we argue that this theory can inform the ongoing efforts to integrate computation into undergraduate physics coursework and programs by highlighting areas of needed development and potential challenges in that process of integration. As previously mentioned, we see a need for frameworks that will provide ways to describe (and assess) computational physics education at a variety of grain sizes, including the student level, classroom/course level, and program level, and we feel that undergraduate physics computational literacy can fill this gap. Third, we see this framework as providing theoretical justification for the uses of computational essays in physics education.

## A. Theoretical implications of the computational literacy framework

The primary theoretical usefulness of this framework is in starting to disentangle the three pillars of computational literacy from one another within the domain of undergraduate physics. As previously mentioned, we see this as a useful theoretical step for the field, because it allows us to distinguish between student issues related to coding experience (material) vs. issues with the process of modeling (cognitive) or presentation (social). We also see this as a natural extension of the ongoing conversation about the necessity of code to teaching computational thinking. If we associate computational thinking with cognitive computational literacy (as have others in the literature, e.g. Berland [35]) then we agree with the prevailing position that computational thinking does not necessitate coding. However, we would contend that even if a physics student is able to think computationally, it does not mean they are computationally literate; computational literacy requires additional engagement with both the material and social aspects of computation.

From a research perspective, we argue that this framework serves those researchers interested in better understanding computational thinking, activity, and literacy broadly, and especially those who are interested in studying the integration of computation and STEM. In particular, we hope this framework will prompt additional conversations within the physics and computer science education research communities about the components of computational theory and practice that are especially relevant to physics. Furthermore, this instantiation of computational literacy for physics raises a number of interesting potential lines of questioning for other STEM fields: What might computational literacy entail for disciplines like mathematics, biology, or chemistry? How do they differ, and why? What types of supports and structures are needed to develop communities who are generally computationally literate, versus those who are computationally literate in particular disciplines like physics? We also feel that this framework might make a useful starting point for larger-scale work on assessing students' computational literacy, for example through widescale deployment of computational essays, combined with the development of a rubric to evaluate different aspects of computational literacy that they demonstrate.

We also wish to highlight the ways students used computationally-generated representations to build their knowledge of physics. This both connects computational literacy to the ever-growing research on student creation and use of representations [57–59] and suggests that this connection could be a productive area of research. It also suggests that any implementation of computation in a physics classroom should be sure to include some kind of visualization tool such as the 3D animations of VPython [60], the 2D animations of Tychos [11], or even simple graphs and plots created by libraries like matplotlib.

## B. Applications of the framework to teaching theory and practice

The primary usefulness of this framework is in applying it to physics teaching theory and practice. First, we see this framework as helping to understand the purpose behind integrating computation into physics teaching and learning. Instructors often talk about how computation allows students to solve more difficult problems, create visualizations, model systems, and "play around" with the physics to discover new things on their own [16,60,61]. However, all of these applications and activities require some degree of computational literacy on the part of the students. Seen from this perspective, computational literacy is a necessary precursor for a student to properly leverage the power of computation in their learning. Based on that argument, we propose that computational literacy is something that should be considered when integrating computation into physics courses. That is, instructors should take their students' levels of computational literacy into account when determining the form and difficulty of computational assignments in their courses.

In the process, we argue that instructors and curriculum designers will also need to pay attention to how students build computational literacy. The University of Oslo has one approach, in which computation is a cornerstone of the physics curriculum and students spend half of their first year familiarizing themselves with programming and mathematical modeling techniques. Other institutions will structure their curricula differently, but we argue that they should be cognizant of which practices, bodies of knowledge, and beliefs they try to instill in their students.

By breaking up computational literacy this way we also want to highlight that different aspects of physics computational literacy will take different timescales to teach, because they are at different levels of abstraction and complexity. Practices can be taught relatively quickly, in that students can be taught the basics of programming and modeling within a few weeks. Knowledge takes longer to develop, because it will require exposure to a variety of tasks and challenges. Beliefs and preferences, we propose, are likely to take even longer to develop. However, as we have argued, such beliefs will likely affect when and how students apply their practices and knowledge. For this reason, beliefs must be explicitly attended to by exposing students to computation in a variety of settings.

We see social computational literacy as an often-neglected pillar within the framework. That is, most computational exercises, projects, and curricula that we are familiar with give students frequent chances to program but do not often ask them to communicate with or about their code to people outside their working group. However, we argue that this skill of communication is extremely important for students to develop.

## C. Usefulness of computational essays in physics teaching

Lastly, we wish to highlight the potential usefulness of computational essays in the physics classroom, and propose this as a case study in that usefulness. Specifically, we argue that these assignments allowed students to both build and demonstrate their computational literacy.

By build, we mean that it gave them an opportunity to combine all three elements of computational literacy in one assignment, thereby practicing the material, cognitive, and social aspects. By demonstrate we mean that it allowed us, as instructors, to see these aspects in practice.

It may seem circular that we have used computational essays as a vehicle for exploring computational literacy, and are now using this framework to argue for their usefulness in the classroom. However, as previously discussed, we chose to use computational essays for this study precisely because we expected them to be fertile ground for students to develop and demonstrate their computational literacy. With this framework articulated, we can revisit that argument and say why, specifically, they are useful: because they allow students to apply their practices, knowledge, and beliefs from the different pillars, and allow instructors to gain insight into the degree of computational literacy their students have achieved. Although we have not yet tried this in practice, we hypothesize that regular use of computational essays, along with regular feedback and guidance on student performance, could help students further develop the various aspects of their computational literacy.

We also see these computational essays as ideal vehicles for open-ended projects, and for building more opportunities for creative exploration into the physics curriculum. Much of physics is highly structured. Aside from certain isolated contexts like open-ended labs [62–65], physics students are seldom given either the tools or the opportunity to "play around" and explore systems on their own. Computation gives them a set of tools that they can use for this kind of exploration, and open-ended projects like the computational essay gives them a framework within which this kind of exploration is supported and encouraged as well as a shareable final form for their project.

In support of these conclusions, it is worth noting that several students in the study reported that they felt very motivated by this open-ended project structure. For example, Lily offered the following reflection in her interview:

> **Lily:** Honestly I think this is better than the obligs [obligatory homework assignments] in a way because I think we pushed ourselves harder here than we would with those assignments. Because then you have an endpoint like, okay, I know what the program or what the assignment asked me to do and here's the program. But now, when we finished something, it was like 'this is really cool to actually see. What else can we do?'

Although computational essays hold great promise, we also see significant challenges with their use. For starters, in the same way that students need training in writing lab reports, students will need guidance in writing computational essays. This may involve both instruction on the structure of a prototypical essay and guidance on the degree to which they should prioritize writing versus coding. There is also a need to develop guidelines for how to effectively grade such essays and where to allocate grade points (for example, on quality of writing, versus code functionality, versus essay structure). Students are also likely to need help in pursuing their open-ended investigations. Because the students in this study volunteered to take part in the computational essay project, we suspect that they likely self-selected for interest in open-ended projects; students who have less interest in these types of projects may need additional supports beyond those provided to our study population.

However, despite these challenges, we see great promise for computational essays in future physics education research and practice.

## VI. CONCLUSION

Nearly 40 years ago, Seymore Papert argued that computation will someday reshape the landscape of teaching and learning, bringing new learning opportunities and allowing students to accomplish things that would have never been possible before. 20 years later, Andrea diSessa proposed his theory of computational literacy. Today, we are beginning to see hints of what a computationally literate generation will mean for society, and for physics.

We see computational essays as a natural step along this path, both in the ways they facilitate physics investigations and in how they bring new opportunities for exploration and communication into the physics curriculum. And we see computational literacy in physics as an outgrowth of this reshaping; that is, it is both a new form of knowledge and a facilitator for deeper physics understanding.

However, more work needs to be done in this area. First, we see a need to continue to develop scaffolds and guidelines around the use of computational essays. We posit that in the years to come, computational essays will emerge as a new genre of scientific communication [66]. However, this is an emerging genre, with few currently-existing guidelines. Second, we need to flesh out this model of physics computational literacy, with data drawn from other contexts and educational levels (from K-12 to professional). Finally, we need more study on how students develop physics computational literacy over time. How do their practices, knowledge, and beliefs evolve, from their first exposure to computation in physics, through their undergraduate years, and into graduate school and beyond? How do different models of computational integration, and different types of assignments, affect this development? And how will students' computational literacy grow and shift as new computational tools and languages emerge? With this framework as a starting point, we look forward to addressing these questions.

## ACKNOWLEDGMENTS

This project was funded by NOKUT, the Norwegian Agency for Quality Assurance in Education, and the Research Council of Norway. We would like to thank Crina Damsa, Christine Lindstrøm, John Burk, Karl Henrik Fredly, and Anders Malthe-Sørenssen for their feedback and help with this study.